\documentclass{appolb-blank}    

\usepackage{graphicx}

\usepackage{amssymb,epsf}
\usepackage{latexsym,,amssymb}
\usepackage{amsmath,mathrsfs}
\usepackage{graphics,epsfig}
\usepackage[english]{babel}

\newcommand{\beq}{\begin{equation}}
\newcommand{\eeq}{\end{equation}}
\newcommand{\beqa}{\begin{eqnarray}}
\newcommand{\eeqa}{\end{eqnarray}}
\newcommand{\bsubeqs}{\begin{subequations}}
\newcommand{\esubeqs}{\end{subequations}}

\begin{document}

\eqsec  

\title{Black-Hole Solution Without Curvature Singularity And\\
Closed Timelike Curves
}
\author{F.R.~Klinkhamer
\address{Institute for
Theoretical Physics, Karlsruhe Institute of
Technology (KIT),\\ 76128 Karlsruhe, Germany}
}
\maketitle
\begin{abstract}
With a prescribed Coulomb-type energy-momentum tensor,
an exact solution of the Einstein field equations over
a nonsimply-connected manifold is presented.
This spherically symmetric solution has neither curvature singularities
nor closed timelike curves.
It can be considered to be a regularization
of the singular Reissner--Nordstr\"{o}m solution
over a simply-connected manifold.\vspace{-2mm}
\end{abstract}

\vspace{-120mm}
\noindent Acta Phys. Pol. B 45, 5 (2014) \hfill arXiv:1305.2875
\vspace{+120mm}

\section{Introduction}
\label{sec:Introduction}

The vacuum Einstein field equations over $\mathcal{M}_4=\mathbb{R}^4$
have a spherically symmetric solution, the Schwarzschild solution
~\cite{Schwarzschild1916,Kruskal1960,Szekeres1960,HawkingEllis1973}.
Recently, a modification of the standard Schwarzschild solution
has been suggested~\cite{Klinkhamer2013-MPLA},
which has no curvature singularity but does have closed timelike curves
inside the Schwarzschild event horizon.
Here, we show that a further  modification
allows us, in principle, to eliminate these closed timelike curves.

The basic idea is as follows. The problematic closed timelike curves
of the modified Schwarzschild solution~\cite{Klinkhamer2013-MPLA} trace
back to the fact that the original singularity
was \emph{spacelike}~\cite{HawkingEllis1973}. But it is well-known
that the singularity of the standard Reissner--Nordstr\"{o}m
solution~\mbox{\cite{Reissner1916,Nordstrom1916,GravesBrill1960, Carter1966}}
is  \emph{timelike}. This suggest, first, to
add a small electric charge and, then, to modify
the resulting Reissner--Nordstr\"{o}m
solution in order to arrive at
a nonsingular black-hole solution without closed timelike curves.

\section{Reissner--Nordstr\"{o}m solution}
\label{sec:RN-BH-Solution}

In this article we use geometric units ($G_N=c=1$)
and consider spherically symmetric solutions
of the Einstein field equations~\cite{HawkingEllis1973},
\bsubeqs\label{eq:Einstein-equations-T-ext}
\beqa\label{eq:Einstein-equations}
R_{\mu}^{\;\;\nu} - \frac{1}{2}\, R\, \delta_{\mu}^{\;\;\nu}
&=&
8\pi\,T_{\mu}^{\;\;\nu} \,,
\eeqa
where the energy-momentum tensor $T_{\mu}^{\;\;\nu}$
is set equal to a prescribed energy-momentum
tensor $\Theta_{\mu}^{\;\;\nu}$ (for spherical coordinates),
\beqa\label{eq:T-ext}
\hspace{-6mm}
T_{\mu}^{\;\;\nu}(t,\,r,\,\theta,\,\phi)
&=&
\Theta_{\mu}^{\;\;\nu}(t,\,r,\,\theta,\,\phi)
\equiv
\frac{Q^2}{8\pi\,r^4}\;
\Big[\text{diag}(-1,\,-1,\,1,\,1)\Big]_{\mu}^{\;\;\nu}.   
\eeqa
\esubeqs
This particular $\Theta_{\mu}^{\;\;\nu}$
corresponds to the energy-momentum tensor
of a Cou\-lomb-type electric field.

The standard Reissner--Nordstr\"{o}m (RN)
solution~\cite{Reissner1916,Nordstrom1916}
has a metric in the exterior region given by the following line element:
\beqa\label{eq:RN-solution}
\hspace*{-0mm}
ds^2\,\Big|^{r>r_{+}}_\text{RN}
&=&
-\left(1- \frac{2 M}{r} + \frac{Q^2}{r^2}\right)\; dt^2
+\left(1- \frac{2 M}{r} + \frac{Q^2}{r^2}\right)^{-1}\; dr^2
\nonumber\\
&&
+ r^2\, \Big( d\theta^2 +\sin^2\theta\; d\phi^2 \Big)\,,
\eeqa
with coordinates $t \in \mathbb{R}$,
$r>r_{+}$, $\theta\in [0,\,\pi]$, and $\phi\in [0,\,2\pi)$,
where $r_{\pm}\equiv M \pm \sqrt{M^2-Q^2}$.
Here, $M$ can be interpreted as the mass of the central object
and $Q$ as its electric charge.

The RN metric in the interior regions ($r\leq r_{+}$)
can best be described with other
coordinates~\cite{HawkingEllis1973,GravesBrill1960, Carter1966}. But, in
the innermost region ($r<r_{-}$), it is possible to revert to $r$ and $t$,
and the metric takes again the form \eqref{eq:RN-solution}.

\section{Nonsingular black-hole solution with electric charge}  

\label{sec:Nonsingular-BH-solution}

\subsection{Topology}
\label{subsec:Topology}

The spacetime considered in this article
corresponds to a noncompact, orientable, nonsimply-connected manifold
$\widetilde{\mathcal{M}}$
without boundary. This manifold  has the topology%
\bsubeqs\label{eq:new-solution-region-III-Mtilde4-Mtilde3}
\beqa\label{eq:new-solution-region-III-Mtilde4}
\widetilde{\mathcal{M}}
 &=& \mathbb{R} \times \widetilde{\mathcal{M}}_{3} \,,
 \\[2mm]
\label{eq:new-solution-region-III-Mtilde3}
\widetilde{\mathcal{M}}_{3}
&\simeq&
\mathbb{R}P^3 - \{\text{point}\}\,,
\eeqa
\esubeqs
where  $\mathbb{R}P^3$ is the 3-dimensional
real projective space (topologically equivalent to
a 3-sphere with antipodal points identified).
The particular manifold $\widetilde{\mathcal{M}}_{3}$
has been discussed extensively
in Refs.~\cite{BernadotteKlinkhamer2006,Schwarz2010,Klinkhamer2013-review},
but the present article aims to be self-contained
and the necessary details will be provided. 

\begin{figure}[t] 
\begin{center}
\includegraphics[width=0.45\textwidth]{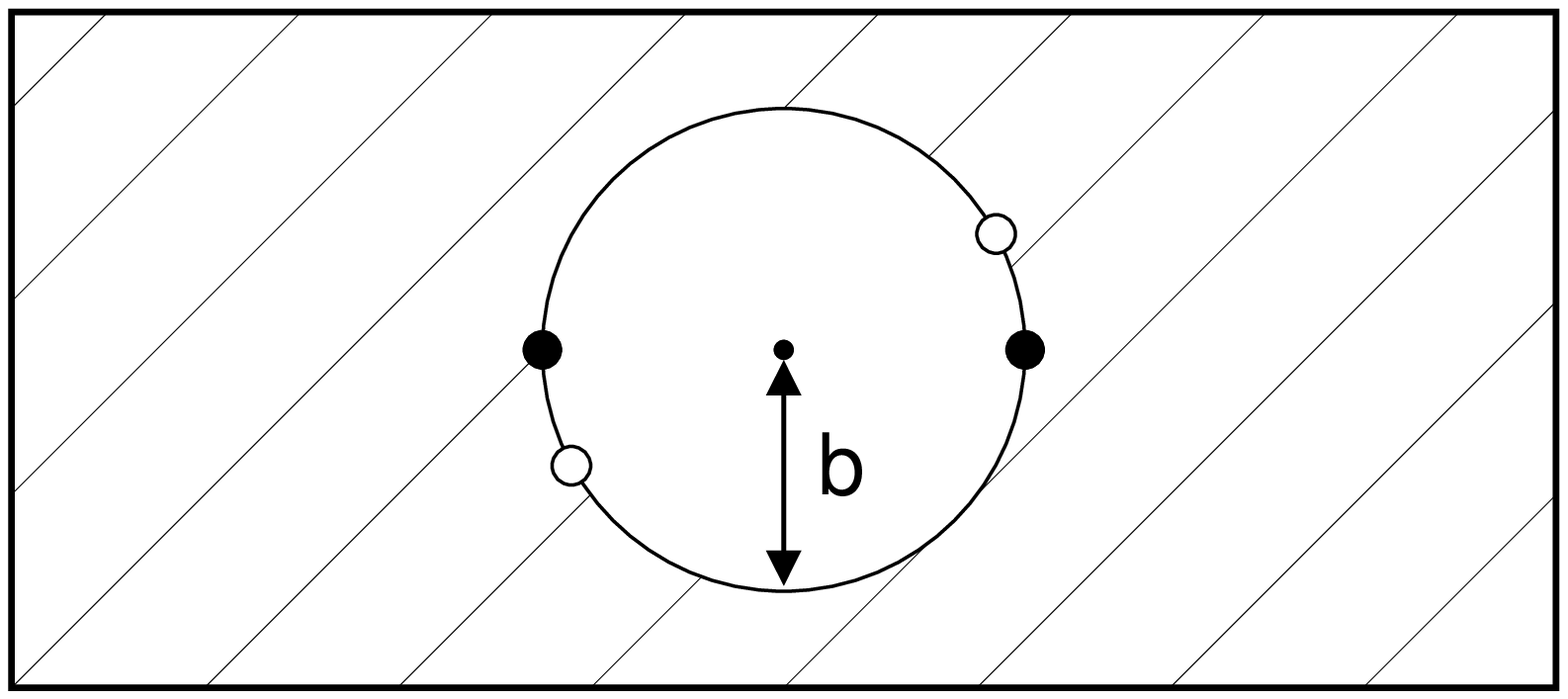}
\end{center}
\caption{Three-space $\widetilde{M}_{3}$
obtained by surgery on $\mathbb{R}^3$:  
interior of the ball with radius $b$ removed
and antipodal points on the boundary of the ball identified 
(as indicated by open and filled circles).}
\label{fig:defect}
\end{figure}

The explicit construction of $\widetilde{\mathcal{M}}_{3}$
is as follows: start from 3-dimensional Euclidean space $E_3$,
remove the interior of a ball ($r<b$),
and identify antipodal points on the boundary ($r=b$).  
See Fig.~\ref{fig:defect} for a sketch and
App.~\ref{app:Manifold} for details.
The \textit{Ansatz} metric will be given in terms of the time coordinate
$T\in \mathbb{R}$ and the proper coordinates of $\widetilde{\mathcal{M}}_{3}$.
Note that the standard Cartesian coordinates of $E_3$ are inappropriate:
different Cartesian coordinates of $E_3$ may correspond to a single point of
$\widetilde{\mathcal{M}}_{3}$ (an example is given by the
filled circles in Fig.~\ref{fig:defect}, which correspond to
a unique point of $\widetilde{\mathcal{M}}_{3}$).

The manifold $\widetilde{\mathcal{M}}_{3}$
is covered by three coordinates charts,
\beqa\label{eq:XnYnZn}
(X_{n},\,Y_{n},\,Z_{n})\,,
\eeqa
for $n=1,2,3$.
These coordinates have the following ranges:
\bsubeqs\label{eq:XnYnZn-ranges}
\beqa\label{eq:X1Y1Z1-ranges}
X_{1} \in (-\infty,\,\infty) \,,\quad
Y_{1} \in (0,\,\pi)\,,\quad
Z_{1} \in (0,\,\pi)\,,
\eeqa
\beqa\label{eq:X2Y2Z2-ranges}
X_{2} \in (0,\,\pi)\,,\quad
Y_{2} \in (-\infty,\,\infty)\,,\quad
Z_{2} \in (0,\,\pi)\,,
\eeqa
\beqa\label{eq:X3Y3Z3-ranges}
X_{3} \in  (0,\,\pi)\,,\quad
Y_{3} \in (0,\,\pi)\,,\quad
Z_{3} \in  (-\infty,\,\infty)\,.
\eeqa
\esubeqs
In each chart, there is one radial-type coordinate with infinite range,
one polar-type angular coordinate of finite range, and one
azimuthal-type angular coordinate of finite range.
Further details can be found in App.~\ref{app:Coordinate-charts}.

\subsection{Parameters and solution}
\label{subsec:Parameters-and-solution}

The nonsingular RN-type solution of the Einstein field equations
has an additional parameter, the length $b$
(an operational definition of $b$ will be given later).
The three parameters of the solution are assumed to be related as follows:
\bsubeqs\label{eq:conditions-QM-b}
\beqa\label{eq:conditions-QM}
0 \;\; < &|Q|&   <  \;\; M\,,
\\[2mm]\label{eq:conditions-b}
0  \;\; < & b & <  \;\; \zeta_{-}\,,
\eeqa
\esubeqs
with definitions
\beqa\label{eq:beta-pm}
\zeta_{\pm} &\equiv& M \pm \sqrt{M^2-Q^2}\,.
\eeqa
Note that, for the classical theory,
the electric charge $|Q|$ can be arbitrarily small, as long as it remains
nonzero.

The construction of the nonsingular solution
with parameters \eqref{eq:conditions-QM-b}
involves an effective radial coordinate $\zeta$,
defined in terms of the quasi-radial coordinates
of $\widetilde{\mathcal{M}}_{3}$ and the length parameter $b$.
We refer to Carter's original article~\cite{Carter1966}
for the conformal structure of the
standard Reissner--Nordstr\"{o}m solution
and follow the same modification procedure as used in
our previous article~\cite{Klinkhamer2013-MPLA} for the Schwarzschild solution.
As we are primarily interested in the removal of the
curvature singularity, we focus on the spacetime
region III ($\zeta < \zeta_{-}$).
No essential changes occur for the spacetime regions I and II
($\zeta \geq \zeta_{-}$), because they do not reach the singularity
(see Fig.~1b of Ref.~\cite{Carter1966}
or Fig.~25 of Ref.~\cite{HawkingEllis1973}).

The construction starts with the chart-1 coordinates as presented in
Sec.~\ref{subsec:Topology}, the other two charts will be added afterwards.
In terms of these chart-1 coordinates $(T,\, X_{1},\, Y_{1},\, Z_{1})$,
the \textit{Ansatz} for the region--III line element is as follows:
\bsubeqs\label{eq:new-solution-region-III-chart-1}
\beqa\label{eq:new-solution-region-III-ds2-chart-1}
\hspace*{-0mm}
ds^2\,\Big|^{b \leq\zeta < \zeta_{-}}_\text{chart-1}
&=&
- \left(1-\frac{2 M}{\zeta}+\frac{Q^2}{\zeta^2}\right)\;dT^2
\nonumber\\&&
+  \left(1-\frac{2 M}{\zeta}+\frac{Q^2}{\zeta^2}\right)^{-1} \;
\frac{(X_{1})^2}{\zeta^2}\;\;(d X_{1})^2
\nonumber\\&&
+ \zeta^2\, \Big( (d Z_{1})^2
+ \big(\sin Z_{1}\big)^2\; (d Y_{1})^2 \Big)\,,
\\[2mm]
\label{eq:new-solution-region-III-zeta-chart-1}
\zeta\,\Big|_\text{chart-1}
&=& \sqrt{b^2+(X_{1})^2}\,.
\eeqa
\esubeqs
As the apparent singularities at
$\zeta=\zeta_{\pm}$ in \eqref{eq:new-solution-region-III-ds2-chart-1}
are away from the ``stitched-up'' surface at $\zeta=b$
(referring to the surgery performed in App.~\ref{app:Manifold}),
the standard analysis~\cite{HawkingEllis1973,Carter1966}
in terms of $T$ and $\zeta$ coordinates
shows that these  apparent singularities
can be removed by appropriate coordinate transformations.

This essentially completes the construction
of our new metric \eqref{eq:new-solution-region-III-chart-1}.
Note that \eqref{eq:new-solution-region-III-ds2-chart-1}
takes precisely the form of the original
Reissner--Nordstr\"{o}m metric \eqref{eq:RN-solution}
if $\big[(X_{1})^2/\zeta^2\big]\,(dX_{1})^2$ is replaced by $d\zeta^2$
according to \eqref{eq:new-solution-region-III-zeta-chart-1}.
But, as emphasized in Ref.~\cite{Klinkhamer2013-MPLA},
the crucial point here is the appearance of
the coordinate $X_{1} \in (-\infty,\,\infty)$
of the nonsimply-connected manifold $\widetilde{\mathcal{M}}_{3}$.
In addition, there are now radial geodesics passing through $X_{1}=0$,
as explained in Sec.~3 of Ref.~\cite{Klinkhamer2013-review}.

The Riemann curvature
tensor $R^{\kappa}_{\;\;\lambda\mu\nu}(T,\,X_{1},\, Y_{1},\, Z_{1})$
from the metric \eqref{eq:new-solution-region-III-chart-1}
is found to be even in $X_{1}$ and finite at $X_{1}=0$.
The Ricci tensor $R_{\mu}^{\;\;\nu}(T,\,X_{1},\, Y_{1},\, Z_{1})$
from \eqref{eq:new-solution-region-III-chart-1}
equals $(Q^2/\zeta^4)\,\text{diag}(1,\,1,\,-1,\,-1)$
and the Ricci scalar $R(T,\,X_{1},\, Y_{1},\, Z_{1})$ vanishes identically.
The metric \eqref{eq:new-solution-region-III-chart-1} solves,
therefore, the Einstein field equations \eqref{eq:Einstein-equations}
for a  prescribed  energy-momentum tensor $\Theta_{\mu}^{\;\;\nu}(T,\,X_{1},\, Y_{1},\, Z_{1})$
of the diagonal form \eqref{eq:T-ext}
with $1/r^4$ replaced by $1/\zeta^4=1/\big(b^2+(X_{1})^2\big)^2$.

The results from App.~\ref{app:Coordinate-charts} allow for an immediate
extension of the metric \eqref{eq:new-solution-region-III-chart-1}
of the $n=1$ chart to the metrics of the $n=2$ and $n=3$ charts:
\bsubeqs\label{eq:new-solution-region-III-chart-2}
\beqa\label{eq:new-solution-region-III-ds2-chart-2}
\hspace*{-0mm}
ds^2\,\Big|^{b \leq\zeta < \zeta_{-}}_\text{chart-2}
&=&
- \left(1-\frac{2 M}{\zeta}+\frac{Q^2}{\zeta^2}\right)\;dT^2
\nonumber\\&&
+  \left(1-\frac{2 M}{\zeta}+\frac{Q^2}{\zeta^2}\right)^{-1} \;
\frac{(Y_{2})^2}{\zeta^2}\;\;(d Y_{2})^2
\nonumber\\&&
+ \zeta^2\, \Big( (dZ_{2})^2 + \big(\sin Z_{2}\big)^2\; (dX_{2})^2 \Big)\,,
\\[2mm]
\label{eq:new-solution-region-III-zeta-chart-2}
\zeta\,\Big|_\text{chart-2}
&=& \sqrt{b^2+(Y_{2})^2}\,,
\eeqa
\esubeqs
and
\bsubeqs\label{eq:new-solution-region-III-chart-3}
\beqa\label{eq:new-solution-region-III-ds2-chart-3}
\hspace*{-0mm}
ds^2\,\Big|^{b \leq\zeta < \zeta_{-}}_\text{chart-3}
&=&
- \left(1-\frac{2 M}{\zeta}+\frac{Q^2}{\zeta^2}\right)\;dT^2
\nonumber\\&&
+  \left(1-\frac{2 M}{\zeta}+\frac{Q^2}{\zeta^2}\right)^{-1} \;
\frac{(Z_{3})^2}{\zeta^2}\;\;(dZ_{3})^2
\nonumber\\&&
+ \zeta^2\,  \Big( (dY_{3})^2 + \big(\sin Y_{3}\big)^2\; (dX_{3})^2 \Big)\,,
\\[2mm]
\label{eq:new-solution-region-III-zeta-chart-3}
\zeta\,\Big|_\text{chart-3}
&=& \sqrt{b^2+(Z_{3})^2}\,.
\eeqa
\esubeqs

The corresponding Kretschmann curvature scalar
over the different charts is given by
\begin{eqnarray}\label{eq:metric-Ansatz-Kscalar}
K  &\equiv& R_{\mu\nu\rho\sigma}\,R^{\mu\nu\rho\sigma}
=\frac{8 \,\big(6\, M^2\, \zeta^2- 12\, M\, Q^2\, \zeta + 7\, Q^4\big)}
      {\zeta^8}\,,
\end{eqnarray}
which remains finite because $\zeta >0$ for  $b>0$.
With fixed values of $M$ and $Q$
obeying condition \eqref{eq:conditions-QM},
$K(\zeta)$ drops monotonically with $\zeta$.
This fact allows for an operational definition of $b$
from the maximum value of $K$. (The operational definitions
of $M$ and $Q$  rely, for example, on the asymptotic $\zeta\to\infty$
behavior of the metric and electromagnetic field.)
Note that the actual value of $b$ sets the length of the
shortest possible noncontractible loop in the
spacelike hypersurface with constant $T$
(such a loop corresponds to half of a great circle
on the sphere $\zeta=b$, taken between antipodal points
which are identified).

The spacetime with metrics
\eqref{eq:new-solution-region-III-chart-1}--\eqref{eq:new-solution-region-III-chart-3}
corresponds
to a noncompact, orientable, nonsimply-connected manifold without boundary
and has the topology \eqref{eq:new-solution-region-III-Mtilde4-Mtilde3}.
The main result of this article is that the
factor $\mathbb{R}$ in \eqref{eq:new-solution-region-III-Mtilde4}
corresponds to the timelike direction of the metrics
\eqref{eq:new-solution-region-III-chart-1}--\eqref{eq:new-solution-region-III-chart-3}, 
making for a spacelike hypersurface $\widetilde{\mathcal{M}}_{3}$
in the spacetime region III. In turn, this observation implies the absence
of closed timelike curves.\footnote{Note that the spacetime regions
I ($\zeta >\zeta_{+}$) and II ($\zeta_{-}< \zeta < \zeta_{+}$)
do not reach the $\zeta=b$ surface where antipodal points are
identified (\textit{cf.} Fig.~\ref{fig:defect}).}

\section{Discussion}
\label{sec:Discussion}

The nonsingular solution
\eqref{eq:new-solution-region-III-chart-1}--\eqref{eq:new-solution-region-III-chart-3}
with parameters \eqref{eq:conditions-QM-b}
provides a ``regularization'' of the singular Reissner--Nordstr\"{o}m
solution. But this regularized spacetime does have a  ``blemish,''
as mentioned in the Note Added of Ref.~\cite{Klinkhamer2013-MPLA}
and detailed in Appendix D of Ref.~\cite{Klinkhamer2013-review}.
The fact is that the coordinate transformation which
brings the manifold
\eqref{eq:new-solution-region-III-chart-1}
near $X_{1}=0$ to a patch of Minkowski
spacetime is a $C^1$ function
with a discontinuous second derivative at $X_{1}=0$ (that is, not a genuine
diffeomorphism, which is a $C^\infty$ function everywhere).  
Whether or not such a classical spacetime without the standard
elementary-flatness property
(having a type of ``spacetime defect'')
plays a role in physics may be up to
quantum gravity to decide, at least according the following
scenario~\cite{Klinkhamer2013-MPLA}.

Start from a nearly flat spacetime
(trivial topology $\mathbb{R}^4$ and
metric approximately equal to the Minkowski metric), where
a large amount of matter with total mass $M$ and with
vanishing net charge $Q=0$
is arranged to collapse in a spherically symmetric way.
Within the realm of classical Einstein gravity, we expect
to end up with the singular Schwarzschild solution.  

But, very close to the final curvature singularity,
something else may happen due to quantum effects.
Considering a precursor mass $\Delta M \sim \hbar/(b\,c) \ll M$
and using typical curvature values
from the expressions for the Kretschmann scalar,
the local spacetime integral of the action density
related to the Schwarzschild solution
differs from that related to \eqref{eq:new-solution-region-III-chart-1}
by an amount $\lesssim \hbar$. Then, as argued by Wheeler
in particular,
the local topology of the manifold may change by a quantum jump
if $b$ is sufficiently close to $L_\text{Planck}\equiv (\hbar\, G_N/c^3)^{1/2}$.
In addition, the strong gravitational fields may lead to electron-positron
pair creation, possibly with one charge expelled towards spatial infinity.

These two quantum processes combined may
result in a transition from a simply-connected
manifold without localized charge to a non\-simply-connec\-ted manifold
with localized charge $Q=\pm\, |Q_\text{electron}|\equiv \pm\, e$.
Hence, if the transition amplitude between the different topologies
is nonzero for appropriate matter content,
quantum mechanics can operate a change between
the classical Schwarzschild solution
and the classical solution \eqref{eq:new-solution-region-III-chart-1} with
$Q=\pm\, e$ and an additional charge $\mp\, e$ at infinity,
thereby removing the curvature singularity
while avoiding closed timelike curves.
The removal of the curvature singularity comes
at the price of introducing a type of spacetime defect
(the blemish mentioned above).
The underlying quantum theory of gravity must
determine if such spacetime defects are allowed or not.

\section*{\hspace*{-4.5mm}Acknowledgments}
\noindent
This work has been supported, in part,
by the ``Helmholtz Alliance for Astroparticle Physics (HAP),''
funded by the Initiative and Networking Fund of the Helmholtz Association.

\begin{appendix}

\section{Manifold}
\label{app:Manifold}

The explicit construction of the 3-space $\widetilde{M}_3$
proceeds by local surgery~\cite{BernadotteKlinkhamer2006}
on the 3-dimensional Euclidean space
$E_3=\big(\mathbb{R}^3,\, \delta_{mn}\big)$.
It is convenient to use standard Cartesian and spherical coordinates,
\beq\label{eq:Cartesian-spherical-coord}
\vec{x}
= (x^1,\,  x^2,\, x^3)
= (r \sin\theta  \cos\phi,\,r \sin\theta  \sin\phi,\, r \cos\theta )\,,
\eeq
with $x^m \in (-\infty,\,+\infty)$,
$r \geq 0$, $\theta \in [0,\,\pi]$, and $\phi \in [0,\,2\pi)$.

Now, $\widetilde{M}_3$ is obtained from $\mathbb{R}^3$
by removal of the interior of the ball $B_b$ with radius $b$ and
identification of antipodal points on the boundary $S_b \equiv \partial B_b$. 
With point reflection denoted $P(\vec{x})=-\vec{x}$,
the 3-space $\widetilde{M}_3$ is given by
\beq\label{eq:M3-definition}
\widetilde{M}_3 =
\Big\{  \vec{x}\in \mathbb{R}^3\,:\; \Big(|\vec{x}| \geq b >0\Big)
\wedge
\Big(P(\vec{x})\cong \vec{x} \;\;\text{for}\;\;  |\vec{x}|=b\Big)
\Big\}\,,
\eeq
where $\cong$ stands for point-wise identification (Fig.~\ref{fig:defect}).

It can be shown that $\widetilde{M}_3$ is a
manifold~\cite{BernadotteKlinkhamer2006,Schwarz2010} and
appropriate coordinate charts will be given in
App.~\ref{app:Coordinate-charts}.
Preparing for that discussion, introduce already
the following nonstandard spherical coordinates
$(r,\, \vartheta,\, \varphi)$  on $\mathbb{R}^3$:
\beqa\label{eq:Cartesian-nonstandard-spherical-coord}
(x^1,\,  x^2,\, x^3)
&=&
(r \sin\vartheta \sin\varphi,\,r \cos\vartheta,\, r \sin\vartheta\cos\varphi)\,,
\eeqa
with
$r \geq 0$, $\vartheta \in [0,\,\pi]$, and $\varphi \in [0,\,2\pi)$.

\section{Coordinate charts}
\label{app:Coordinate-charts}

\begin{figure}[b] 
\begin{center}
\includegraphics[width=0.6\textwidth]{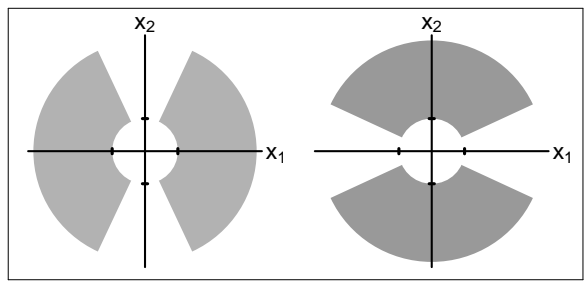}
\end{center}
\caption{Slice $x_3 = 0$ of the manifold $\widetilde{\mathcal{M}}_3$ with the
domains of the chart-1 coordinates (left) and
the chart-2 coordinates (right). The tick marks on the
$x_1$ and $x_2$ axes correspond to the values $\pm b$
(see Fig.~\ref{fig:defect}). The 3-dimensional
domains are obtained by revolution
around the $x_1$-axis (left) or the $x_2$-axis (right).
The domain of the chart-3 coordinates is defined similarly.}
\label{fig:defect-charts}
\vspace{0cm}
\end{figure}

The 3-space  $\widetilde{M}_3$ was defined in
App.~\ref{app:Manifold} and shown in Fig.~\ref{fig:defect}.
A relatively simple covering~\cite{Schwarz2010} of $\widetilde{M}_3$
uses three charts of coordinates, labeled by $n=1,2,3$.
Each  chart covers and surrounds part of one of the three Cartesian
coordinate axes but does not intersect the other two Cartesian coordinate axes.
For example, the $n=1$ coordinate chart covers and surrounds
the $|x^1|\geq b$ segments of
the $x^1$ coordinate axis but does not intersect the $x^2$ and $x^3$  axes.
The domains of the chart-1 coordinates consist of two `wedges,'
on both sides of the defect
and pierced by the $x^1$ axis;  see Fig.~\ref{fig:defect-charts}.

These coordinates are denoted
$(X_n,\,  Y_n,\, Z_n)$, for  $n=1,\,2,\,3$.
Note that, despite appearances, the triples $(X_n,\,  Y_n,\, Z_n)$
are non-Cartesian coordinates.

Referring to the standard spherical
coordinates \eqref{eq:Cartesian-spherical-coord}
of the Euclidean 3-space $E_3$, the chart-1
coordinates over the relevant regions of $\widetilde{M}_{3}$
(\textit{i.e.}, the wedges of Fig.~\ref{fig:defect-charts}, left) 
are given by
\bsubeqs\label{eq:X1Y1Z1-def}
\beqa
X_{1} &=&
\left\{\begin{array}{ll}
r-b  \hspace*{7.5mm} &\quad\text{for}\quad  \cos\phi > 0\,,\\
b-r       &\quad\text{for}\quad  \cos\phi <    0\,,
\end{array}\right.\\[2mm]
Y_{1} &=&
\left\{\begin{array}{ll}
\phi-\pi/2  &\quad\text{for}\quad  \pi/2 < \phi< 3\pi/2\,,\\
\phi-3\pi/2 &\quad\text{for}\quad  3\pi/2 < \phi< 2\pi\,,\\
\phi+\pi/2  &\quad\text{for}\quad  0 \leq \phi < \pi/2\,,
\end{array}\right.\\[2mm]
Z_{1} &=& \left\{\begin{array}{ll}
\theta       &\quad\text{for}\quad  \cos\phi > 0\,,\\
\pi-\theta\hspace*{7.0mm}   &\quad\text{for}\quad  \cos\phi <    0\,,
              \end{array}\right.
\eeqa
with ranges
\beqa\label{eq:X1Y1Z1-ranges-appendix}
X_{1} \in (-\infty,\,\infty) \,,\quad
Y_{1} \in (0,\,\pi)\,,\quad
Z_{1} \in (0,\,\pi)\,.
\eeqa
\esubeqs

The construction of the chart-2 coordinates is entirely analogous
to those of the chart $n=1$.
Specifically, this set of coordinates over the relevant regions
(wedges of Fig.~\ref{fig:defect-charts}, right) 
of $\widetilde{M}_{3}$ is given by
\bsubeqs\label{eq:X2Y2Z2-def}
\beqa
X_2 &=& \left\{\begin{array}{ll}
\phi       &\quad\text{for}\quad  0 < \phi< \pi\,,\\
\phi-\pi   &\quad\text{for}\quad  \pi < \phi < 2\pi\,,
              \end{array}\right.\\[2mm]
Y_2 &=& \left\{\begin{array}{ll}
r-b   \hspace*{1.0mm}    &\quad\text{for}\quad  0 < \phi< \pi\,,\\
b-r       &\quad\text{for}\quad  \pi < \phi < 2\pi\,,
              \end{array}\right.\\[2mm]
Z_2 &=& \left\{\begin{array}{ll}
\theta       &\quad\text{for}\quad  0 < \phi< \pi\,,\\
\pi-\theta   &\quad\text{for}\quad  \pi < \phi < 2\pi\,.
              \end{array}\right.
\eeqa
with ranges
\beqa\label{eq:X2Y2Z2-ranges-appendix}
X_2 \in (0,\,\pi)\,,\quad
Y_2 \in (-\infty,\,\infty)\,,\quad
Z_2 \in (0,\,\pi)\,.
\eeqa
\esubeqs

For the $n=3$ chart, we require nonstandard spherical coordinates
that are regular on the Cartesian $x^3$ axis.
These have been defined in \eqref{eq:Cartesian-nonstandard-spherical-coord}.
Now, the chart-3 coordinates over the relevant regions (wedges)
of $\widetilde{M}_{3}$ are given by
\bsubeqs\label{eq:X3Y3Z3-def}
\beqa
X_{3} &=&
\left\{\begin{array}{ll}
                \varphi-\pi/2  &\quad\text{for}\quad  \pi/2 < \varphi< 3\pi/2\,,\\
                \varphi-3\pi/2 &\quad\text{for}\quad  3\pi/2 < \varphi< 2\pi\,,\\
                \varphi+\pi/2  &\quad\text{for}\quad  0 \leq \varphi < \pi/2\,,
\end{array}\right.\\[2mm]
Y_{3} &=&
\left\{\begin{array}{ll}
\vartheta       &\quad\text{for}\quad  \cos\varphi > 0\,,\\
\pi-\vartheta \hspace*{7.0mm}   &\quad\text{for}\quad  \cos\varphi <    0\,,
\end{array}\right.  \\[2mm]
Z_{3} &=&
\left\{\begin{array}{ll}
r-b  \hspace*{7.5mm}     &\quad\text{for}\quad  \cos\varphi > 0\,,\\
b-r       &\quad\text{for}\quad  \cos\varphi <    0\,,
\end{array}\right.
\eeqa
with ranges
\beqa\label{eq:X3Y3Z3-ranges-appendix}
X_{3} \in  (0,\,\pi)\,,\quad
Y_{3} \in (0,\,\pi)\,,\quad
Z_{3} \in  (-\infty,\,\infty)\,.
\eeqa
\esubeqs

Having expressed the coordinates $(X_n,\,  Y_n,\, Z_n)$  in terms of
coordinates of the Euclidean 3-space,
it is possible to verify that the $(X_n,\,  Y_n,\, Z_n)$ coordinates
are invertible and infinitely-differentiable functions of each other
in the overlap regions. These coordinates therefore describe a manifold.
Moreover, the manifold satisfies~\cite{Schwarz2010} the Hausdorff property
(two distinct points $x$ and $y$ are always surrounded by two disjoint
open sets $U$ and $V$: $x\in U$, $y\in V$, and $U\cap V= \emptyset$).

\end{appendix}


\end{document}